\documentclass[prb,twocolumn,superscriptaddress,showpacs,epsf]{revtex4}
\usepackage{rotating}
\usepackage{graphicx}
\usepackage{latexsym}
\usepackage{amssymb}
\usepackage{amsfonts}
\usepackage{soul}
\usepackage{color}
\begin{document}
\title{Self organized mode locking effect in superconductor / ferromagnet hybrids}

\author{J. Van de Vondel}
\affiliation{INPAC -- Institute for Nanoscale Physics and Chemistry, Nanoscale Superconductivity \\ and Magnetism Group, K.U.Leuven, Celestijnenlaan 200D, B--3001 Leuven, Belgium}
\author{A. V. Silhanek}
\affiliation{INPAC -- Institute for Nanoscale Physics and Chemistry, Nanoscale Superconductivity \\ and Magnetism Group, K.U.Leuven, Celestijnenlaan 200D, B--3001 Leuven, Belgium}
\author{ V. Metlushko}
\affiliation{Department of Electrical and Computer Engineering, University of Illinois, Chicago, Illinois 60607, USA}
\author{ P. Vavassori}
\affiliation{Physics Department, University of Ferrara and INFM-National Research Center on nanoStructures and Biosystems at Surfaces (S3), via Saragat 1, I-44100 Ferrara, Italy} \affiliation{CIC nanoGUNE Consolider, Mikeletegi Pasealekua 56, 301, E-20009 - Donostia-San Sebastian, Spain}
\author{ B. Ilic}
\affiliation{Cornell Nanofabrication Facility, School of Applied and Engineering Physics, Cornell University, Ithaca, New York 14853, USA}
\author{V. V. Moshchalkov}
\affiliation{INPAC -- Institute for Nanoscale Physics and Chemistry, Nanoscale Superconductivity \\ and Magnetism Group, K.U.Leuven, Celestijnenlaan 200D, B--3001 Leuven, Belgium} 

\date{\today}
\begin{abstract}

The vortex dynamics in a low temperature superconductor deposited
on top of a rectangular array of micrometer size permalloy
triangles is investigated experimentally. The rectangular unit
cell is such that neighboring triangles physically touch each
other along one direction. This design stabilizes remanent states
which differ from the magnetic vortex state typical of individual
non-interacting triangles. Magnetic Force Microscopy images have
revealed that the magnetic landscape of the template can be
switched to an ordered configuration after magnetizing the sample
with an in-plane field. The ordered phase exhibits a broad flux
flow regime with relatively low critical current and a highly
anisotropic response. This behavior is caused by the spontaneous
formation of two separated rows of vortices and antivortices along
each line of connected triangles. The existence of a clear flux
flow regime even for zero external field supports this
interpretation. The density of induced vortex-antivortex pairs is
directly obtained using a high frequency measurement technique
which allows us to resolve the discrete motion of vortices.
Strikingly, the presence of vortex-antivortex rows gives rise to a
self organized synchronized motion of vortices which manifests
itself as field independent Shapiro steps in the current-voltage
characteristics.

\end{abstract}


\maketitle

\section{introduction}

The dynamics and pinning of vortices in type-II superconductors is
arguably one of the most widely investigated phenomena in the
field of superconductivity. Recently, particular attention has
been devoted to the possibility of controlling the strength of the
vortex pinning using different magnetic
templates\cite{magneticpinning, Gillijns07}. Within the London
formalism, the interaction between a vortex line and a finite size
permanent magnet is given by\cite{Milosevic-PRB-03b},

\begin{equation}
    U_{p}({\bf R})=-\int_{dot} {\bf m}({\bf r})\cdot{\bf B}_v({\bf R-r})~d^3r, \label{Uvm-dot}
    \end{equation}

\noindent where the integration is carried out over the volume of
the ferromagnet, ${\bf R}$ indicates the position of the vortex
line, ${\bf m}({\bf r})$ its spatial dependent magnetic moment,
and ${\bf B}_v$ is the field generated by the vortex line. This
equation shows that the pinning potential not only depends on the
size of the dots\cite{Lyuksyutov-AdvPhys-05,Gillijns07} but also
on their exact magnetic state\cite{Gheorghe-PRB-08}. In
particular, if the superconducting penetration depth $\lambda$ is
on the order of the lateral size of the dot, a weak average
pinning is expected when the magnetic dot contains multiple
domains. In contrast to that, a maximum pinning should be achieved
for single domain structures. In order to retain these properties
in big arrays of magnetic particles it is necessary to ensure a
minimum distance between them in such a way that the field
profiles of neighboring elements do not overlap and can be
resolved at scales of the magnetic penetration length $\lambda$.

Notice that according to Eq.(\ref{Uvm-dot}) it is possible to
increase strongly the pinning by just growing continuously the
magnetic moment of the dots. This growth is limited by the fact
that above certain magnetic moment, a vortex-antivortex pair will
be induced by the magnetic element with the consequent change in
the pinning properties\cite{Milosevic-PRB-04}. For instance, dots
with perpendicular magnetic moment induce a vortex on top of the
dots whereas the accompanying antivortex is located at
interstitial positions, in between the dots. As a result, vortices
and antivortices feel a different pinning potential which leads to
a distinct dynamic response\cite{Lange-PRB-05}. Since the weakly
pinned interstitial antivortices limit the maximum critical
current $J^{max}_c$, their annihilation with an external field
produces a shift of the field position of the $J^{max}_c$ towards
$H=nH_1$, where $n$ is the number of vortex-antivortex pairs
generated by a single magnetic element and $H_1$ is the field
corresponding to one external vortex per unit cell. Since $H_1$ is
a geometrically predefined parameter, the field loci of the
$J^{max}_c$ is a direct measurement of the number $n$ of bounded
pairs. Unfortunately, in the case of micromagnets with in-plane
magnetic moment, the field polarity symmetry of the system
conceals this information from dc electro-transport experiments.

In this work, we introduce alternative methods to accurately
determine the number $n$ of vortex-antivortex pairs induced by the
magnetic template, based on the ac-dynamic response of the hybrid
system. Indeed, the fact that the critical force to bring the
vortex and antivortex closer differs from that necessary to pull
them apart\cite{Carneiro-PhysC-05}, gives rise to a clear
rectification effect when the system is submitted to a zero
average ac-signal. This rectified voltage is maximum at zero
external field\cite{Clecio07} and progressively decreases with
increasing field until all vortices (or antivortices) are
compensated.  At RF frequencies the mode locking or synchronized
motion of vortices produces distinct voltage steps in the
current-voltage characteristics with separation determined by the
number of moving vortices. Both techniques indicate that for the
particular system here investigated, the magnetic template
generates one single vortex-antivortex pair per unit cell.

The crucial conditions for the success of these methods are (i) a
periodic landscape, i.e. all vortices and antivortices feeling the
same environment, and (ii) a free flux flow motion of vortices.
Although the predefined topological landscape is by definition
periodic, the magnetic landscape for the as-grown state consists
of a highly disorder distribution of magnetic poles. This
disordered state can be switched to a perfectly ordered magnetic
lattice by magnetizing the sample with an in-plane field. The
order-disorder transition, already reported for other S/F
systems\cite{Silhanek-APL-07}, produces a profound modification in
the vortex dynamics. More specifically, in the disordered state,
non-linear current-voltage characteristics with high critical
currents are observed, whereas a very broad flux flow regime
accompanied by a sharp decrease of the critical currents is
obtained in the ordered state. This behavior can be attributed to
the reduction of the average hopping distance determined by the
separation between neighboring magnetic poles of identical
polarity. In addition, the presence of a flux flow regime at zero
applied field reinforces the idea that the investigated magnetic
structure is able to induce vortex-antivortex (v-av) pairs.


 \section{Sample preparation and characterization}
\begin{figure}[b!]

\includegraphics[width=0.3\textwidth,angle=-0]{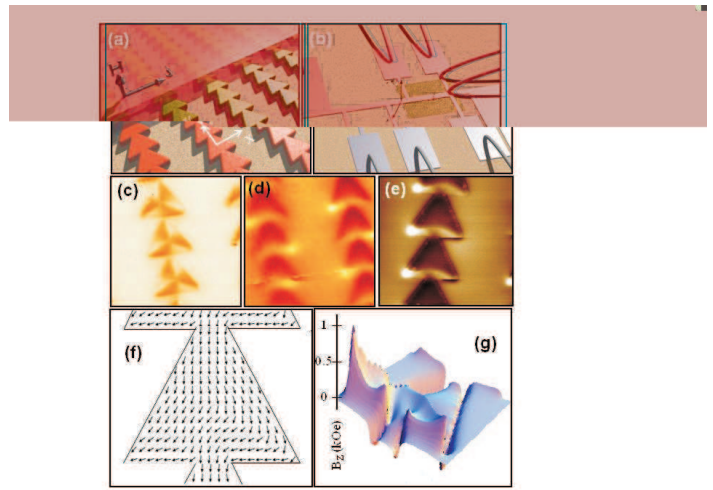}
\caption{(color online)(a) schematic drawing of the array of Py
triangles. The direction of the external field, the current, and
the relative orientation of the array is clearly indicated with
arrows. (b) Schematic representation of the Al transport bridge
with Hall voltage contacts on the patterned region. Magnetic Force
Microscopy pictures obtained at remanence of physically separated
magnetic triangles (c), touching magnetic triangles in the virgin
state (d), and the triangles magnetized perpendicular to the chain
of triangles (e). Panels (f) and (g) show the local magnetic
moment distribution and the z-component of the stray field,
respectively, calculated by OOMMF simulations
\cite{micromagnetic}.} \label{Fig-Sample}
\end{figure}
The sample investigated consists of a periodic array of permalloy
(Ni$_{80}$Fe$_{20}$) equilateral triangles of 1 $\mu$m side and a
thickness of 25 nm. Along one of the principal directions of the
array ($x$-axis) the separation between the triangles is 1 $\mu$m
whereas in the perpendicular direction ($y$-axis) the triangles
are touching [see Fig.~\ref{Fig-Sample}(a)]. This geometry gives
rise to two well defined directions of the structure, parallel
($y$) and perpendicular ($x$) to the line connecting nearby
triangles, as indicated in Fig.~\ref{Fig-Sample}(a). A 50 nm thick
superconducting Al bridge deposited on top of the magnetic
template also covers the unpatterned substrate [see
Fig.~\ref{Fig-Sample}(b)] thus allowing us to compare directly
patterned vs reference plain film and identify the influence of
the magnetic triangles on the superconductor. Both, the array of
triangles and the transport bridge are aligned at submicrometer
scale and were prepared by standard lithographic techniques. The
Al plain film has a critical temperature at zero field $T_{c0} =
1.315$ K, with residual resistivity ratio $R(300$K$)/R(4.2$K$)
\sim 2$, and superconducting coherence length $\xi$(0)$ \approx
128$ nm.

It is a crucial point that neighboring triangles overlap in one
direction in order to eventually achieve two well distinguished
phases: disordered in the as-grown state and ordered in the
magnetized state. Indeed, Magnetic Force Microscopy (MFM) images
shown in Fig.~\ref{Fig-Sample}(c) for non touching triangles
demonstrate that the remanent state, irrespective of the magnetic
history of the triangles, is always a vortex state, in agreement
with previous reports on individual micrometer size magnetic
structures\cite{vortex, Miltat-02}. In contrast to that, if there
is a physical contact between neighboring triangles a new set of
possible states can be found. Fig.~\ref{Fig-Sample}(d) shows a MFM
image of the sample investigated in the as-grown state. Here a
clearly disordered arrangement of the, so called, buckle
states\cite{koltsov} is observed. Interestingly, after magnetizing
the sample with an homogeneous in-plane field perpendicular to the
chain of triangles [Fig.~\ref{Fig-Sample}(e)], these buckle states
can be set in a perfectly ordered arrangement. The in-plane
distribution of magnetic moments calculated by OOMMF micromagnetic
simulations\cite{micromagnetic} is indicated in
Fig.~\ref{Fig-Sample}(f) with black arrows whereas the
$z$-component of the stray field ($B_z$) is shown in
Fig.~\ref{Fig-Sample}(g). In all cases the magnetic state of each
triangle is characterized by a strong out of plane magnetic pole
which is located at one of the two bottom corners of the Py
triangles and a weaker magnetic pole, with opposite polarity,
located at the top corner of the triangle
[Fig.~\ref{Fig-Sample}(e)]. It is precisely this magnetic pole
which can be controlled to lie on either side of the triangle by
simply changing the direction of magnetization. Notice that this
magnetic monopole-like structure breaks the field polarity
symmetry of the S/F hybrid system. The question now arises as
whether the distribution of the magnetic dipoles (ordered versus
disordered) has any effect on the properties of the adjacent
superconducting film.

\section{Nucleation of superconductivity}

\begin{figure}[h!]
\includegraphics[width=0.45\textwidth]{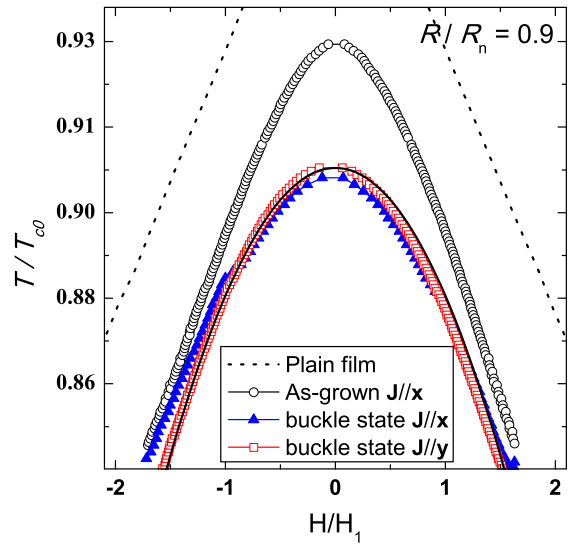}
\caption{Superconducting/normal Phase boundary as determined by 90
$\%$ resistance criterion of the normal resistance $R_{n}$ and
dc-current of 100 $\mu$A for the as-grown state (open circles) and
different magnetized states and current orientations. The dashed
line shows the corresponding transition lines for the
co-evaporated Al plain film. The solid line is a fit assuming a
lateral confinement of the superconducting condensate (see text).}
\label{Fig-Phaseboundary}
\end{figure}

Let us first analyze the nucleation of the superconducting order
parameter for the two different magnetic states of the hybrid
system. Fig.~\ref{Fig-Phaseboundary} summarizes the obtained
superconductor/normal metal phase boundaries for a dc-current of
100 $\mu$A and using 90$\%$ criterion of the normal state
resistance $R_{n}$ for different magnetic states and current
orientations. Higher criteria exhibit similar results. For the
as-grown state (open circles) a pronounced decrease of the
critical temperature at low fields with respect to the plain film
is observed. A further suppression of $T_c$ is obtained after the
sample has been magnetized with an in-plane field parallel to the
$x$ direction (solid triangles). The fact that the phase boundary
of the as-grown state lies in between those corresponding to the
plain film and the fully polarized buckle state, suggests that in
the virgin sample, randomly oriented buckle states coexist with
magnetic vortex states. This has been indeed confirmed by the MFM
images taken at different locations of the pattern.


It is worth noticing that for the magnetized sample the $T_c(H)$
boundary exhibits a parabolic background similar to that reported
in earlier investigations\cite{Lange-2003b,Rosseel-1997,
VVM-1995}. This behavior can be ascribed to the change of
dimensionality when $\xi(T)$ exceeds the size $w$ of the regions
where superconductivity first nucleates. Within this regime the
phase boundary can be approximated by $T_c(H)/T_c(0)=1-(\alpha
H)^2$, with $\alpha=\xi(0)\pi w /\sqrt 3 \phi_0$. By using this
expression to fit the data in Fig.~\ref{Fig-Phaseboundary} (solid
line) we found $w \sim 1.4$ $\mu$m which is consistent with the
largest separation between neighboring magnetic poles. This
behavior suggests that for the magnetized sample the stray fields
confine the superconducting condensate in between the magnetic
poles and consequently the superconducting order parameter will be
modulated into a sequence of quasi one-dimensional stripes. This
prediction is in agreement with complementary transport
measurements for two orthogonal orientations of the applied
current, perpendicular to the chain of triangles (solid triangular
symbols) and parallel to them (open triangular symbols) as shown
in Fig.~\ref{Fig-Phaseboundary} indicating that the onset of
superconductivity is independent on the direction of the applied
current.

\section{Vortex dynamics in the ordered and disordered phases}

In the previous section we showed that by switching from the
as-grown (disordered phase) to the magnetized (ordered phase)
state, a proliferation of magnetic poles takes place which in turn
gives rise to a decrease of $T_{c}$. In addition, the dimensional
crossover from a 2D disordered phase to a 1D ordered phase has
little or no influence on the nucleation temperature. However,
deeper into the superconducting state where the electric response
is dominated by the vortex dynamics, the one dimensional nature of
the field modulation should become more apparent.

\begin{figure}[h!]
\includegraphics[width=0.45\textwidth]{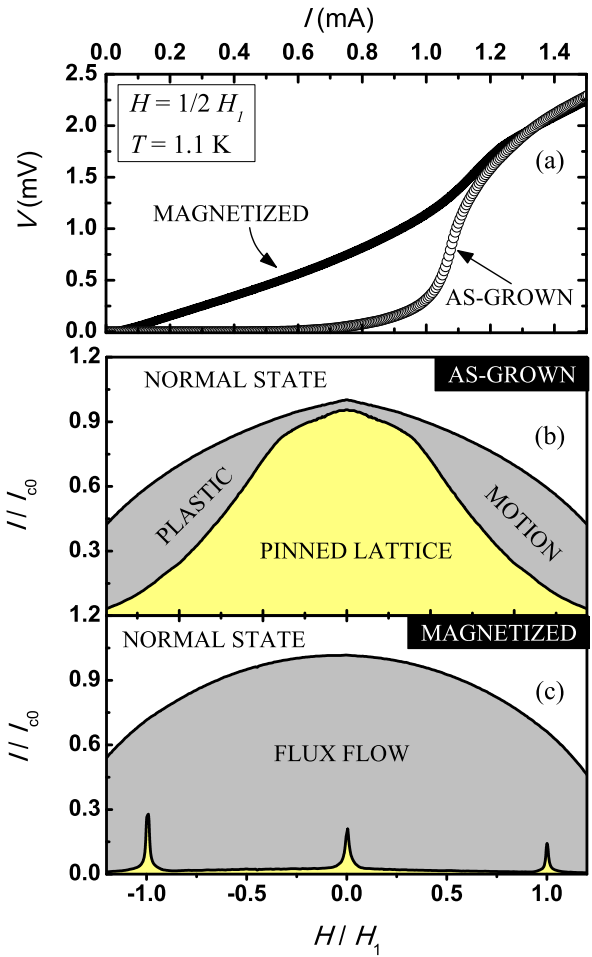}
\caption{(color online)(a) Voltage-current $V(I)$ characteristic
with current applied in the $x$-direction, at $H =
\frac{1}{2}H_{1}$ and a temperature of 1.1 K for the as-grown and
the magnetized states. The lower panels show the magnetic field
dependence of the critical current and the transition to the
normal state as determined by criterion of 10$\%$ R$_{n}$ and
90$\%$ R$_{n}$, respectively,  for the as-grown (b) and magnetized
(c) states.} \label{VI}
\end{figure}

In order to address this issue we have measured the
voltage-current characteristics [$V(I)$] with the bias current
applied in the $x$-direction at $H = \frac{1}{2}H_{1}$ and $T =
1.1$ K for the two different magnetic states: as-grown (open
circles) and magnetized (solid squares) states as shown in
Fig.~\ref{VI} (a). The most obvious feature in this Figure is the
enormous difference (more than an order of magnitude) between the
critical current of the magnetized state (24 $\mu$A) with respect
to the demagnetized state (490 $\mu$A). Furthermore, the vortex
dynamics is severely modified after magnetizing the sample.
Indeed, whereas in the as-grown state there is a narrow window of
currents exhibiting a highly non-linear $V(I)$ dependence, where
the vortex lattice transit from pinned phase to the normal state,
in the magnetized sample a broad flux-flow regime is observed
immediately above the critical current. Fig.~\ref{VI}(b) and (c)
summarize the field dependence of the different dynamical phases
for the as grown sample [panel (b)] and after magnetization [panel
(c)]. The onset of vortex motion is delimited by a resistance
criterion of 10$\%$ R$_{n}$ while a 90$\%$ R$_{n}$ criterion is
used to indicate the transition to the normal state.

Notice that the presence of matching features, which require
two-dimensional phase coherence, together with the fact that in
the magnetized state a finite critical current is needed to depin
the vortices, rule out the possibility of having normal metal
stripes in the hybrid sample. Instead, the observed difference
between both phases can be attributed to the formation of easily
moving channels of vortices and antivortices created by the
magnetic structure. Indeed, in the ordered phase the potential
wells created by the magnetic dipoles are perfectly aligned and
strongly overlap, thus creating an easy channel for flux motion.
In contrast to that, in the disordered phase, the average distance
between two neighboring magnetic pinning centers of equal polarity
is larger thus impeding the easy motion of vortices. From now on
we will focus on the dynamic response of the system in the ordered
state.

\section{Evidence of induced v-av pairs}

A rough criterion to determine whether the stray field emanating
from the magnetic structure is strong enough to induce v-av pairs
can be obtained by comparing it to the penetration field $H_p$
resulting from Bean-Livingston barriers\cite{Melnikov98}. The
highest barrier possible is limited by the thermodynamic critical
field $H_p=\Phi_0/(2\sqrt{2} \pi \lambda \xi)$. Using $\xi=128$ nm
and $\lambda=150$ nm we obtain $H_p \sim 12$ mT which lies well
below the field intensity ($\sim$ 100 mT) produced by the magnetic
template at the superconducting film. Based on this estimation we
can safely conclude that the magnetic structure is able to induce,
at least, one vortex-antivortex per unit cell.

 \begin{figure}[hbt]
\includegraphics[width=0.45\textwidth]{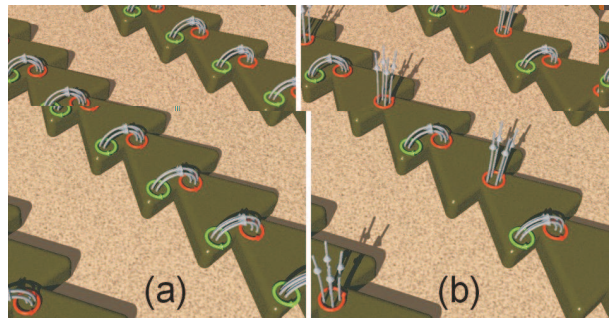}
\caption{(color online) Schematic drawing of the induced vortex
lattice in the case of no external applied field (a) and an
applied field equal to $\frac{1}{2}H_{1}$ (b). Red (green) circles
represent positive (negative) vortices. } \label{Schematics}
\end{figure}

\begin{figure}[t]
\includegraphics[width=0.45\textwidth]{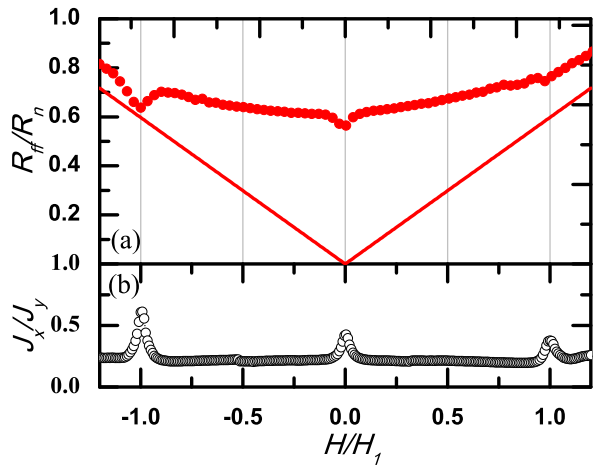}
\caption{ (a) The field dependence of the flux flow resistance for
the magnetized sample at $T$ = 1.1 K. The straight lines indicate
the expected dependence according to free flux flow model. (b)
Critical current as a function of magnetic field at a temperature
of 1.1 K for both current directions. A clear anisotropic response
is observed.} \label{Comparison}
\end{figure}

This finding suggests that the ordered phase consists of two well
defined channels with opposite vortex polarity as schematically
shown in Fig.~\ref{Schematics}(a). Since the magnetic template
breaks the field polarity symmetry, in principle vortices and
antivortices should experience a different depinning current.
Moreover, as shown in Fig.~\ref{Schematics}(b), under these
circumstances an external field will compensate vortices in the
channel with opposite polarity (thus introducing disorder in this
particular channel and increasing the average hopping distance)
and as a result the high vortex mobility will be suppressed. In
contrast to that, the row with vortices of opposite polarity would
remain unaltered. This scenario predicts two anomalous behaviors:
the existence of a flux-flow response at zero field and a field
independent flux flow regime as long as none of the two rows of
vortices is fully compensated by the external field. Both
predictions are in agreement with the field evolution of the
$V(I)$ measurements done at 1.1 K. The slope of the linear regime,
proportional to the amount of particles contributing to the easy
flow, is plotted as a function of field in
Fig.~\ref{Comparison}(a). In comparison to the slope expected for
the free flux flow model $R_{ff}=R_n H/H_{c2}$ [straight line in
Fig.~\ref{Comparison}(a)], where $R_n$ is the normal state
resistance and $H_{c2}$ is the upper critical field, a weak field
dependence of the flux flow slope is seen at low field values.

Further evidence in favor of the existence of channels populated
with v-av pairs comes from the measurements of the critical
current for currents applied parallel ($J_y$) and perpendicular
($J_x$) to the channels of v-av pairs. These results are shown in
Fig.~\ref{Comparison}(b), with a critical current determined using
a criterion of 100 $\mu$V. This figure shows a strong in-plane
anisotropy $\gamma=J_x/J_y$ induced by the magnetic template which
results from the easier motion of vortices along the line of
magnetic poles than perpendicular to them. This anisotropy amounts
to a factor of 5 at non matching fields and is minimum at the
commensurability fields.

\section{Determination of the v-av pairs density at low frequencies: ratchets}

The scenario described in the previous section where rows of
vortices and antivortices generated by the magnetic template
coexist, makes our system similar to the array of elongated
magnetic bars studied by de Souza Silva {\it et
al.}\cite{Clecio07}. This system has been modeled within the
London approximation by Carneiro\cite{Carneiro-PhysC-05} for the
case where no v-av pairs are induced, and more recently by Lima
and de Souza Silva\cite{Cond-mat08} assuming that v-av pairs
emanate from the dipoles. Irrespective of whether the magnetic
template is able to induce v-av pairs or not, both situations lead
to a rectification of the net vortex motion $v$ when a zero
average ac-force is applied parallel to the dipolar moment (i.e.
along the line connecting a vortex with its corresponding
antivortex)\cite{Rectification}. However, clear differences in the
field dependence of the average velocity are expected. Indeed, if
v-av pairs are present then (i) a maximum rectification is
expected at zero field, (ii) unlike standard ratchet signal, the
sign of the average voltage readout should be field polarity
independent, (iii) as vortices (antivortices) are progressively
compensated by increasing the external negative (positive) field,
the rectified signal should decrease.

 \begin{figure}[hbt]
\includegraphics[width=0.45\textwidth]{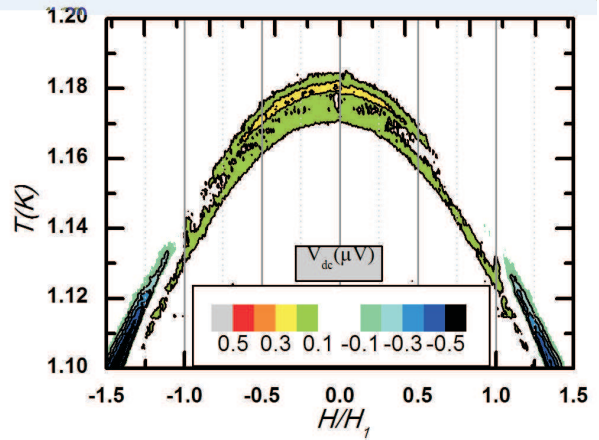}
\caption{(color online) Rectified voltage $V_{dc}$ in the
temperature-field $(T-H)$ plane for the magnetized sample
subjected to a 1 kHz sinusoidal ac current in the $y$ direction of
amplitude 100 $\mu$A.} \label{Ratchet}
\end{figure}

The measured rectification effect, shown in Fig.\ref{Ratchet},
confirms all the above mentioned facts and indicates that the v-av
pairs are more easily brought together than separated.
Furthermore, if the applied field is increased above the first
matching field a sign reversal occurs. This sign reversal at $H_1$
can not be explained assuming more than one v-av pair created by
the magnetic template since in this case the v-av annihilation
process would be unlikely interrupted by the partial compensation
of one row of vortices. In contrast to that, if there is a single
v-av pair per unit cell, once a whole line is compensated by the
external field, a dramatic change in the rectification properties
is expected. Although this finding seems to indicate that the
magnetic structure is able to induce a single v-av per triangular
element, a more convincing evidence of this prediction can be
obtained only by monitoring the individual motion of vortices at
higher frequencies.

 \section{Determination of the v-av pair density at high frequencies: mode locking}

The motion of vortices in a periodic potential gives rise to a
variety of interesting dynamical effects such as the generation of
rf radiation\cite{Martinoli75} and synchronization/resonance
effects of the vortex lattice\cite{Fiory71,Daldini74, lieve99}
when a rf current $I_{rf}$ is superimposed onto a dc drive
$I_{dc}$. The latter effect gives rise to a series of plateaus in
the $V(I)$ measurements very similar to the Shapiro steps observed
in Joshepson junctions\cite{Shapiro63}. The relevant timescale of
these processes is determined by the washboard frequency, $f =
v/d$ with $d$ the period of the potential and $v$ the average
velocity of the particles. The investigation of the above
mentioned phenomena can be used to obtain additional information
on the dynamics of the vortex lattice\cite{Kokubo08}, the
periodicity of the underlying potential and the density of
particles in the system\cite{Kokubo02}.  In other words, high
frequency experiments provide further insight into the microscopic
motion of vortices not available with the averaged response
presented in previous sections.

The synchronous motion of vortices resulting from the resonant
condition $f = v/d$, known as mode locking effect, allows one to
obtain the exact number of v-av pairs created by the magnetic
template. Indeed, the value of the average velocity is given by:
\begin{equation} v = cfd
\end{equation}

\noindent with $d$ the periodicity of the lattice in the direction
of motion, $f$ the frequency of the rf current and $c$ the number
of cells that a vortex has traveled during a rf cycle. As we
mentioned above, the fingerprint of this type of resonant motion
is the presence of clear plateaus in the $I(V)$ characteristics
separated by voltage steps,

\begin{equation}
V_{c}=Nk \frac{h}{2e}\frac{v}{d}=Nk \frac{h}{2e}cf ,
\label{discrete_steps}
\end{equation}

\noindent with $k$ the number of vortices in each unit cell and
$N$ the number of rows between two voltage contacts. Measuring the
voltage separation between consequtive steps and using
Eq.(\ref{discrete_steps}) we can estimate the density of vortices
moving coherently in the flux flow regime.

\begin{figure}[hbt]
\includegraphics[width=0.45\textwidth]{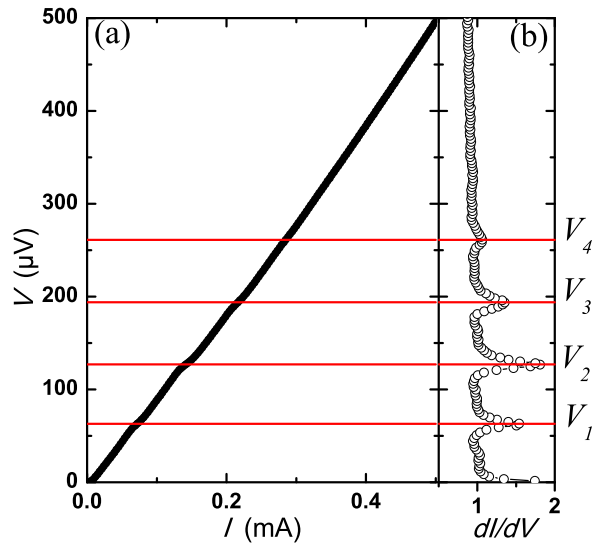}
\caption{(a) $V(I)$ measurement (filled symbols) at $H$ = 0.5 mT
and $T$ = 1.1 K with a small ac excitation of 100 MHz. Panel (b)
shows a $dI/dV$, the discrete peaks are labeled according to
Eq.(\ref{discrete_steps}).} \label{Single measurement shapiro}
\end{figure}

Fig.~\ref{Single measurement shapiro}(a) shows a $V(I)$ scan for
$H=$0.5 mT and $T=$1.1 K with a small ac excitation
$I_{rf}<<I_{dc}$ of 100 MHz superimposed. The presence of Shapiro
steps in the linear part of the $V(I)$ curve becomes more apparent
by plotting $V$ versus $dI/dV$ [see Fig.~\ref{Single measurement
shapiro}(b)]. These steps represent a clear indication of the
coherent flow of vortices.

\begin{figure}[t]
\includegraphics[width=0.45\textwidth]{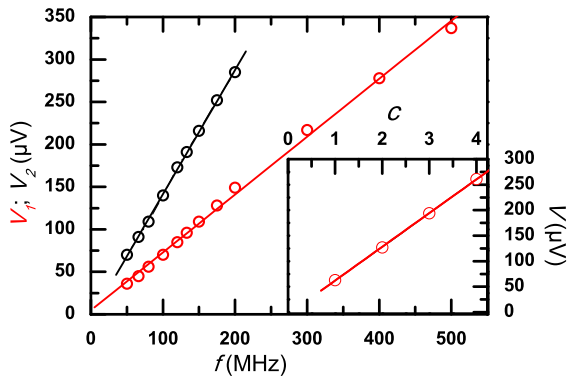}
\caption{Frequency dependence of $V_{1}$ (circular symbols) and
$V_{2}$ (square symbols) at a field of 0.5 mT and temperature of
1.1 K. Inset: $V_{1} \rightarrow V_{4}$ measured at 100 MHz.}
\label{Frequency dependence}
\end{figure}

In Fig.~\ref{Single measurement shapiro}(b) we assign the present
peaks to values of $V_c$ with $c$ ranging from 1 to 4 ($V_{1} = 63
\mu V$, $V_{2} = 127 \mu V$, $V_{3} = 194 \mu V$ and $V_{4} = 261
\mu V$). From their linear dependence on $c$, shown in the inset
of Fig.~\ref{Frequency dependence}, we calculate a value of $k
\approx 1.06$. An independent estimation can be obtained by
tracking the frequency dependence of each voltage plateau $V_{c}$.
This is shown in Fig.~\ref{Frequency dependence} for the first two
steps. The linear fit using Eq.(\ref{discrete_steps}) gives $k
\approx 1.1$ and 1.18 for $V_{1}$ and $V_{2}$, respectively.

It is important to notice that, in general, two ingredients are
necessary for the presence of mode locking effects. Firstly, the
existence of a flux flow regime, which ensures a narrow
distribution of average vortex velocities\cite{Martinoli76}.
Secondly, there should be a periodic perturbation of the pinning
potential, otherwise the distortion of the vortex lattice becomes
too strong and its coherent motion is destroyed. In principle,
this second condition is only fulfilled at the matching field,
$H_{1} = 1.195$ mT. This seems to be in contradiction with the
fact that clear Shapiro steps are present at 0.5 mT, a field which
would corresponds to 0.42 vortices per unit cell. This apparent
paradox can be solved invoking once again the presence of v-av
pairs induced by the triangles. In this case, the applied field
will annihilate the vortices in the channel with opposite
polarity, leading to an increase of their critical current and
consequently suppressing their contribution to the coherent
motion. On the other hand, the row with equal polarity will be
unaltered by the applied magnetic field and therefore should
exhibit a region of field independent Shapiro steps. In other
words, the incoming vortices generated by the external magnetic
field play no role in the coherent oscillatory motion of vortices
thus giving rise to a constant voltage separation between two
consecutive Shapiro steps.

\begin{figure}[t]
\includegraphics[width=0.45\textwidth]{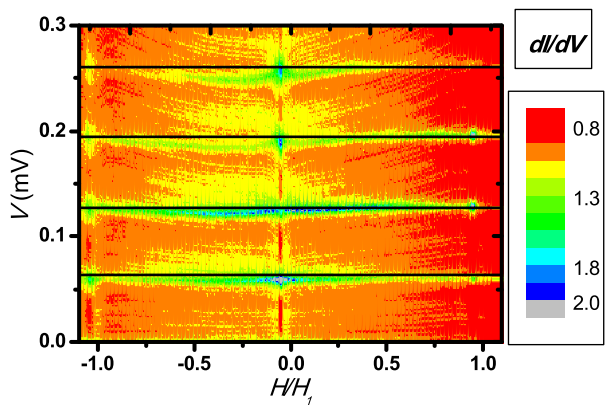}
\caption{(color online) Magnetic field dependence of $dI/dV$
versus $V$ at a temperature of 1.1 K and an applied rf frequency
of 100 MHz.} \label{Magnetic field dependence}
\end{figure}

In order to check this hypotheses of self organized
commensurability we measured the field dependence of the Shapiro
steps, which are better visualized as a peak in the $dI/dV$ as
previously shown in Fig.~\ref{Single measurement shapiro}(b). The
results of these measurements for $T=$1.1 K are summarized in Fig.
~\ref{Magnetic field dependence} as a contour plot in the
field-voltage plane, with maximum intensity representing the exact
location of the $V_c$ peaks. For magnetic field values ranging
from $-H_1$ to $+H_1$ a clear field independence peak position is
observed. This result confirms that the amount of particles,
moving coherently, is independent of the applied external field.

\section{Discrete motion of the vortex-antivortex lattice}

It has been predicted by molecular dynamics simulations that a
vortex lattice in a periodic landscape of asymmetric pinning sites
can unveil the discrete hopping of vortices when the system is
excited with a zero mean rf drive\cite{Janko99,Zhu04,Lu07}. So
far, most of the experimental reports evidence a vortex
rectification effect obtained in the low frequencies limit which
results from an averaging in time and are not able to resolve the
individual motion\cite{Villegas04, VandeVondel05}. In this
section, we will try to bridge this gap by constructing an
artificial vortex rectifier resulting from a dc tilt, which breaks
the inversion symmetry of our system\cite{Clecio06}. Additionally,
this tilt allows us to probe the discrete motion of the vortex
lattice without the need of a time resolved signal read out.

\begin{figure}[hbt]
\includegraphics[width=0.45\textwidth]{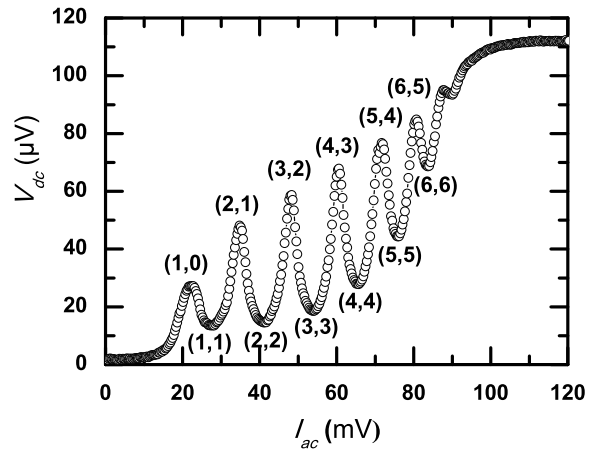}
\caption{DC voltage as a function of ac amplitude ($f$ = 100 MHz)
with a dc offset of 10 $\mu$A applied. The indexing ($a,b$)
indicates the number of steps forward $(a)$ and backward
(b).}\label{Signle_ac_sweep}
\end{figure}

Fig. ~\ref{Signle_ac_sweep} shows the dc voltage as a function of
the applied ac drive with an additional dc current on top
($I_{dc}$ = 10 $\mu$A). Due to the applied dc tilt the critical
current, needed to depin vortices, is smaller for positive than
for negative currents ($I_{pos} < I_{neg}$). Depending on the
strength of the applied ac amplitude $I_{ac}$ we can determine
three different dynamic regimes. For small ac excitations
($I_{ac}<I_{pos} < I_{neg}$) the driving force is unable to depin
the vortices and as a result the dc voltage is zero. By increasing
the ac amplitude above $I_{pos}$ the vortex lattice is set in the
flux flow regime and mode locking effects result in an oscillatory
dependence of the dc voltage as a function of the ac amplitude.
Finally at high ac amplitudes the normal state is reached and
$V_{dc} = R_{n}I_{dc}$. Each maxima and minima in dc voltage can
be attributed to the discrete motion of the vortex lattice in the
positive and negative cycle of the driving amplitude. To identify
the motion, we introduce the indexing ($a,b$) in
Fig.~\ref{Signle_ac_sweep}, with $a$ and $b$ the number of steps
in the positive and negative direction respectively. At $I_{ac}$
just above $I_{pos}$ the driving force is strong enough to move
the vortex lattice one unit cell in the positive direction while
no motion occurs in the negative cycle of the ac drive and a
maxima in the dc voltage output occurs. If the ac amplitude passes
$I_{neg}$ a single back and forth step of the vortex lattice
occurs and as such a strong reduction of the dc voltage is
measured. This  difference between back and forth motion of the
vortex lattice, due to a small dc tilt, results in a discrete
stepping procedure up to the sixth order.

\begin{figure}[hbt]
\includegraphics[width=0.45\textwidth]{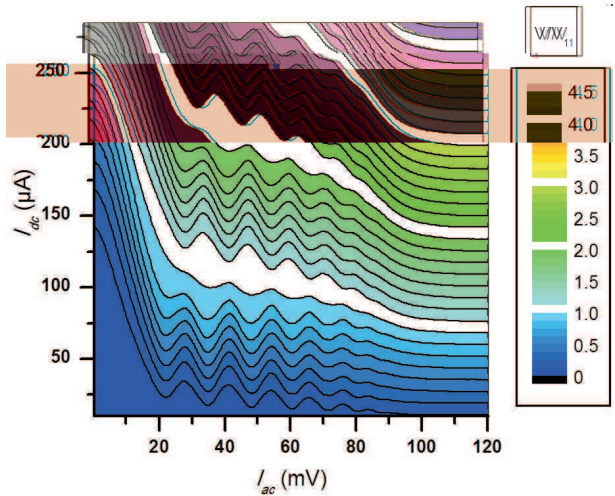}
\caption{(color online) $I_{ac}$ and $I_{dc}$ dependence of the dc
voltage at a magnetic field value of 1.195 mT and a rf modulation
of 150 MHz} \label{Iac_Idc_dependence}
\end{figure}

Fig.~\ref{Iac_Idc_dependence} shows the dependence of the dc
voltage, normalized to $V_{1}$, as a function of the ac drive and
the dc tilt. At small dc tilts (below $\approx$ 100 $\mu$A) the
behavior is exactly the same as shown in
Fig.~\ref{Signle_ac_sweep}. In this region, the applied dc current
creates a small asymmetry and the vortex lattice, depending on the
exact value of the ac drive, can either move one step extra in the
positive direction (maxima in dc voltage) or an equal amount of
steps in both directions (minima in dc voltage). Around 100 $\mu$A
the oscillations vanish (white region) and the dc voltage readout
amounts $V_{1}$. The reason for this behavior is a subtle balance
between the ac drive and the dc tilt. Exactly at the point were
the ac amplitude is strong enough to move the vortex lattice in
the 'hard' direction (i.e. against the dc tilt) it is also
sufficiently strong to move the vortex lattice an additional step
in the `easy' direction. Therefore the overall motion of the
vortex lattice will be independent on the applied ac drive and
equal to 1. This subtle balance is destroyed in favor of `easy'
flow if the applied dc current is increased above 100 $\mu$A.
Again the dc voltage readout ($1<V_{dc}/V_{1}<2$) indicates that
the vortex lattice moves two steps (maxima in dc voltage) or one
step (minima in dc voltage) extra in the positive direction. By
increasing the dc tilt even further we could even see a difference
of 4 unit cells between the motion in the positive and negative
direction.

\section{Conclusion}

We have demonstrated that the particular geometry of the two-fold
symmetric array of magnetic triangles can induce an order-disorder
transition as the system switches from the as-grown state to the
magnetized state. The disordered phase is characterized by a
random distribution of two opposite chirality magnetic buckle
states, whereas in the order state all triangles are in a well
defined unique magnetic buckle state. The latter phase exhibits
unidimensional lines of strong stray fields which represent
channels for easy flux motion resulting in a clear anisotropy of
the critical current. A quantitative estimation shows that the
current needed to induce inter-channels hopping is about 5 times
larger than that needed for setting intra-channel motion. It is
important to stress that even though we have presented clear
evidence of the influence of the ordering of magnetic dipoles
induced by the magnetic templating on the vortex dynamics, it is
still uncertain whether the ultimate source of disorder in the
as-grown state is the random distribution of magnetic poles in the
buckle state or the presence of triangles in the magnetic vortex
state along one line of connecting triangles. In addition, we
showed that the high mobility of the vortex lattice is a
consequence of the generation of vortex-antivortex pairs by the
magnetic template. This picture was confirmed experimentally by
high frequency transport measurements and low frequency ratchet
experiments. Due to the self organized commensurability of these
channels a pronounced mode locking effect is observed. These
features were used to probe and tune the discrete motion of the
composite lattice of vortices and antivortices. This combination
of techniques represent a powerful tool for understanding the
motion of vortices at microscopic scale.

\section{Acknowledgements}

This work was supported by the K.U.Leuven Research Fund
GOA/2004/02 program, NES--ESF program, the Belgian IAP, the Fund
for Scientific Research -- Flanders (F.W.O.--Vlaanderen). V.M.
acknowledge funding support from U.S. NSF, Grant ECCS-0823813
(VM). A.V.S. and J.V.d.V. are grateful for the support from the
FWO-Vlaanderen. Finally the authors also would like to acknowledge
help of ScienTec and Nanotec with the MFM images.


\begin{references}

\bibitem{magneticpinning} A.V. Silhanek, W. Gillijns, V.V. Moshchalkov, V. Metlushko, B. Ilic, Appl. Phys. Lett. {\bf 89}, 182505 (2006); N. Verellen, A. V. Silhanek, W. Gillijns, V.V. Moshchalkov, V. Metlushko, F. Gozzini, and B. Ilic, Appl. Phys. Lett. {\bf 93}, 022507 (2008); A. Hoffmann, L. Fumagalli, N. Jahedi, J. C. Sautner, J. E. Pearson, G. Mihajlovic, V. Metlushko, Phys. Rev. B {\bf 77}, 060506 (2008); J. E. Villegas, K. D. Smith, Lei Huang, Yimei Zhu, R. Morales, I. K. Schuller, Phys. Rev. B {\bf 77}, 134510 (2008); V. Vlasko-Vlasov, U. Welp, G. Karapetrov, V. Novosad, D. Rosenmann, M. Iavarone, A. Belkin, and W.-K. Kwok, Phys. Rev. B {\bf 77}, 134518 (2008); A. Belkin, V. Novosad, M. Iavarone, J. Pearson, and G. Karapetrov, Phys. Rev. B {\bf 77}, 180506(R) (2008).

\bibitem{Gillijns07} W. Gillijns, M. V. Milo\v{s}evic, A.V. Silhanek, V.V. Moshchalkov and F. M. Peeters, Phys. Rev. B {\bf 76}, 184516 (2007).

\bibitem{Milosevic-PRB-03b} M. V. Milo\v{s}evic and F. M. Peeters, Phys. Rev. B {\bf 68}, 094510 (2003).

\bibitem{Lyuksyutov-AdvPhys-05} I. F. Lyuksyutov and V. L. Pokrovsky,  Adv. Phys. {\bf 54}, 67 (2005).

\bibitem{Gheorghe-PRB-08} D. G. Gheorghe, R. J. Wijngaarden, W. Gillijns, A. V. Silhanek, and V. V. Moshchalkov, Phys. Rev. B {\bf 77}, 054502 (2008); W. Gillijns, A.V. Silhanek, V.V. Moshchalkov, Phys. Rev. B {\bf 74}, 220509(R) (2006).

\bibitem{Milosevic-PRB-04} M.V. Milo\v{s}evi\'{c}, F.M. Peeters, Phys. Rev. B {\bf 69}, 104522 (2004).

\bibitem{Lange-PRB-05} M. Lange, M.J. Van Bael, A.V. Silhanek, V.V. Moshchalkov, Phys. Rev. B {\bf 72}, 052507 (2005).

\bibitem{Carneiro-PhysC-05} G. Carneiro, Physica C {\bf 432}, 206 (2005).


\bibitem{Clecio07} C. C. de Souza Silva, A. V. Silhanek, J. Van de Vondel, W. Gillijns, V. Metlushko, B. Ilic, and V. V. Moshchalkov, Phys. Rev. Lett. {\bf 98}, 117005
(2007).

\bibitem{Silhanek-APL-07} A. V. Silhanek, W. Gillijns, V.V. Moshchalkov, V. Metlushko, F. Gozzini, B. Ilic, W. Uhlig, J. Unguris, Appl. Phys. Lett. {\bf 90}, 182501 (2007).

\bibitem{vortex} see C.A.F. Vaz, C. Athanasiou, J. A. C. Bland, G. Rowlands, Phys. Rev. B {\bf 73}, 054411 (2006) and references therein.

\bibitem{Miltat-02} J. Miltat and A. Thiaville, Science {\bf 298}, 5593 (2002).

\bibitem{koltsov} D. K. Koltsov, R. P. Cowburn, and M. E. Welland, J. Appl. Phys. {\bf 88}, 5315 (2000); D. K. Koltsov and M. E. Welland, J. Appl. Phys. {\bf 94}, 3457 (2003).

\bibitem{micromagnetic} Micromagnetic simulation is performed by a public available code from NIST (http://math.nist.gov/oommf).

\bibitem{Lange-2003b} M. Lange, M.J. Van Bael, V.V. Moshchalkov, Phys. Rev. B {\bf 68}, 174522 (2003).

\bibitem{Rosseel-1997} E. Rosseel, T. Puig, M. Baert, M. J. Van Bael, V. V. Moshchalkov and Y. Bruynseraede , Physica C {\bf 282-287}, 1567 (1997).

\bibitem{VVM-1995} V. V. Moshchalkov, L. Gielen, C. Strunk, R. Jonckheere, X. Qiu, C. Van Haesendonck and Y.
Bruynseraede, Nature {\bf 373}, 319 (1995).

\bibitem{Melnikov98} A. S. Mel'nikov, Yu. N. Nozdrin, I. D. Tokman, and P. P.
Vysheslavtsev, Phys. Rev. B {\bf 58}, 11672 (1998).

\bibitem{Cond-mat08} C.L.S. Lima and C.C. de Souza Silva,
arXiv:0808.2421

\bibitem{Rectification} A much weaker rectification signal is
observed when the Lorentz force is parallel to the v-av channels.

\bibitem{Martinoli75} P.Martinoli, O. Daldini, C. Leemann, and E. Stocker, Solid State Commun. {\bf 17}, 205 (1975).

\bibitem{Fiory71} A.T. Fiory, Phys. Rev. Lett. {\bf 27}, 501 (1971).

\bibitem{Daldini74} O. Daldini, P.Martinoli, J.L. Olson and G. Berner, Phys. Rev. Lett. {\bf 32}, 218 (1974).

\bibitem{lieve99} L. Van Look, E. Rosseel, M.J. Van Bael, K. Temst, V. V. Moshchalkov and Y. Bruynseraede, Phys. Rev. B {\bf 60}, R6998 (1999).

\bibitem{Shapiro63} S. Shapiro, Phys. Rev. Lett. {\bf 11}, 80 (1963).

\bibitem{Kokubo08} N. Kokubo, B. Shinozaki, P.H. Kes,  Physica C {\bf 468}, 581 (2008).

\bibitem{Kokubo02} N. Kokubo, R. Besseling, V.M. Vinokur and P.H. Kes, Phys. Rev. Lett. {\bf 88}, 247004 (2002).

\bibitem{Martinoli76} P. Martinoli, O. Daldini, C. Leemann and B. Van den Brandt, Phys. Rev. Lett. {\bf 36} 382 (1976).

\bibitem{Janko99} C.-S. Lee, B. Jank\'{o}, I. Der\'{e}nyi and A.-L. Barab\'{a}si, Nature {\bf 400}, 337 (1999).


\bibitem{Zhu04} B. Y. Zhu, F. Marchesoni, V. V. Moshchalkov, and Franco Nori, Phys. Rev. B {\bf 68}, 014514 (2003).

\bibitem{Lu07} Qiming Lu,C. J. Olson Reichhardt, C. Reichhardt, Phys. Rev. B {\bf 75}, 054502 (2007).

\bibitem{Villegas04} J. E. Villegas, Sergey Savel'ev, Franco Nori, E. M. Gonzalez, J. V. Anguita, R. Garc\'{i}a and J. L. Vicent, Science {\bf 302}, 1188 (2003).

\bibitem{VandeVondel05} J. Van de Vondel, C. C. de Souza Silva, B. Y. Zhu, M. Morelle and V. V. Moshchalkov, Phys. Rev. Lett. {\bf 94}, 057003 (2005).


\bibitem{Clecio06} C. C. de Souza Silva, J. Van de Vondel, B. Y. Zhu, M. Morelle, and V. V. Moshchalkov, Phys. Rev. B {\bf 73}, 014507 (2006).


\end{references}
\end{document}